\documentclass[pre,onecolumn,showpacs,superscriptaddress,groupedaddress]{revtex4}
\usepackage[paperwidth=210mm,paperheight=297mm,centering,hmargin=2cm,vmargin=2.5cm]{geometry}
\usepackage{times}
\usepackage{epsfig}
\usepackage{amsmath,amssymb}

\begin{document}
\title{{Super-Planckian thermal emission from a hyperlens}}
\author{C. Simovski}
\affiliation{Department of Radio Science and Engineering, Aalto University, P.O. Box 13000, FI-00076, Aalto, Finland}
\affiliation{Laboratory of Metamaterials, University for Information Technology, Mechanics and Optics (ITMO), St. Petersburg 197101, Russia}
\author{S. Maslovski}
\affiliation{Departamento de Engenharia Electrot\'{e}cnica, Instituto
  de Telecomunica\c{c}\~{o}es, Universidade de Coimbra, P\'{o}lo II,
  3030-290 Coimbra, Portugal}
\affiliation{Laboratory of Metamaterials, University for Information Technology, Mechanics and Optics (ITMO), St. Petersburg 197101, Russia}
\author{S. Tretyakov}
\affiliation{Department of Radio Science and Engineering, Aalto University, P.O. Box 13000, FI-00076, Aalto, Finland}
\author{I. Nefedov}
\affiliation{Department of Radio Science and Engineering, Aalto University, P.O. Box 13000, FI-00076, Aalto, Finland}
\author{S. Kosulnikov}
\affiliation{Laboratory of Metamaterials, University for Information Technology, Mechanics and Optics (ITMO), St. Petersburg 197101, Russia}
\author{P. Belov}
\affiliation{Laboratory of Metamaterials, University for Information Technology, Mechanics and Optics (ITMO), St. Petersburg 197101, Russia}

%%% Abstract
\begin{abstract}
%%% Start here with text of abstract
  We suggest and theoretically explore a possibility to strongly
  enhance the steady thermal radiation of a small thermal emitter
  using an infrared hyperlens. The hyperbolic metamaterial of the
  hyperlens converts emitter's near fields into the propagating waves
  which are efficiently irradiated from the hyperlens surface. Thus,
  with the hyperlens, emitter's spectral radiance goes well beyond the
  black-body limit for the same emitter in free space.  Although the
  hyperlens can be kept at a much lower temperature than the emitter,
  the whole structure may radiate, in principle, as efficiently as a
  black body with the same size as that of the hyperlens and the same
  temperature as that of the emitter.  We believe that this study can
  lead to a breakthrough in radiative cooling at microscale, which is
  crucial for microlasers and microthermophotovoltaic systems.
\end{abstract}


\maketitle

%\section{Introduction}

In the classical theory of thermal radiation the power radiated from a
unit surface of an optically large body in free space per unit
interval of frequencies is given by Planck's formula:
\begin{equation}
P^{\rm
  FS}_{\omega}=\pi e_s(\omega) B_{\omega}={\hbar\omega^3 e_s(\omega)\over 4\pi^2
  c^2}\left[e^{\hbar\omega/(k_BT)}-1\right]^{-1}, \label{eq:Planck}
\end{equation} where $k_B$ and $\hbar$ are Boltzmann's and Planck's constants,
respectively, $T$ is the temperature of the body surface and
$e_s(\omega)<1$ is the spectral emissivity of body's material. For a black
body (BB) emitting, in accordance to the Planckian theory, maximal
thermal radiation to free space, $e_s(\omega)\equiv 1$. For a material
having no optical losses at frequency $\omega$, i.e. for transparent
media, $e_s(\omega)=0$ and the emission is absent. It is commonly accepted
that the thermal radiation is non-coherent and its spatial
distribution is isotropic. Both these factors result in Lambertian
pattern for a radiating half-space. However, recent investigations
have shown that thermal radiation can be partially coherent \cite{11},
directive \cite{11_1}, and combining coherence and directionality
\cite{111,111_1}. These deviations originate from intrinsic properties
of metamaterials \cite{1111}. Especially, the so-called hyperbolic
metamaterial (HMM) shows interesting responses to thermal radiation
(see e.g. in \cite{1,2,22,222}). In this paper we theoretically reveal
a possibility for a sample of HMM to strongly enhance the far-field
radiation from small (several micrometers) emitters exceeding the BB
limit defined for the emitters of the same size in free space. To our
knowledge, in previous works related with applications of HMMs for
radiative heat transfer these materials were used only to control
near-field thermal flows. Here, we use these media to enhance
far-field radiation.

Since classical works by Kirchhoff and Planck, the BB has been
considered as a perfect thermal emitter whose spectral radiance cannot
be exceeded in far-field zone (so-called Planckian limit).  Despite
that the photonic density of states (PDOS) and, consequently, the rate
of spontaneous emission responsible for the thermal radiation may be
enhanced considerably (e.g. in \cite{Bush} by one order of magnitude)
there is a belief that the photons that occupy the extra available
states cannot be emitted out of a medium with high PDOS~\cite{Trupke}
due to the total internal reflection (TIR).  However, it is not
generally true.

\begin{figure}[b!]
  \centering \vskip-2mm\epsfig{file=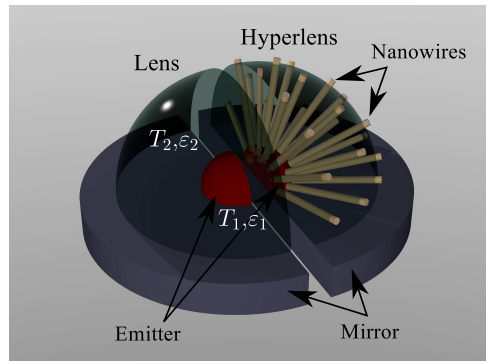, width=0.37\textwidth}
  \caption{\label{fig1}Two hemispherical structures which offer
    super-Planckian thermal radiation from a finite emitter (just a
    half of each structure is shown). Left: Thermal lens analogous to
    the one considered in Ref.~\cite{4}. Right: Thermal hyperlens (HL)
    comprising radially diverging nanowires. The lens and HL's host
    are made of transparent glass with permittivity
    $\varepsilon_2\equiv\varepsilon_{\rm h}$. The emitter is formed by a lossy medium
    with complex permittivity $\varepsilon_1$ and is partially filled with
    nanowires. The emitter is set under high temperature $T_1\gg
    T_2$, where $T_2$ is the ambient temperature.}
\end{figure}

Long ago, in work \cite{3}, a possibility to exceed the BB limit for a
hot particle having resonant sizes at infrared was pointed
out. Recently, the authors of \cite{4} have demonstrated
super-Planckian radiation from a macroscopic emitter achieved due to a
transparent dielectric dome. For a hemispherical emitter this idea is
illustrated in Fig.~\ref{fig1}~(left half). As it follows from
Eq.~(\ref{eq:Planck}), filling free space with an isotropic transparent
medium with refractive index $n=\sqrt{\varepsilon_{\rm h}}$ [respectively, $c$
is replaced by $c/n$ in Eq.~(\ref{eq:Planck})] increases the radiated power
by $n^2$, provided that $e_s(\omega)$ stays unchanged.  If such
transparent medium forms a lens in a shape of a hemispherical dome as
is shown in Fig.~\ref{fig1}, the emitted waves impinge on the lens
surface and, after being partially reflected, pass onto free space.
When the dome radius $R$ is much larger than the emitter radius $r$,
the power transmittance to the free space for these waves can be found
in geometric optics (GO) approximation as for normally incident rays:
$t_{\rm GO}=4n/(n+1)^2$. Thus, the total gain in the power irradiated
to the far zone due to the presence of the dome equals $G=n^2\times
t_{\rm GO} = 4n^3/(n+1)^2$. From here it may seem that one can achieve
arbitrary high gain when $n\rightarrow\infty$. However, this is not
true, because when $n \gtrsim R/r$ some of the incident rays start
experiencing TIR at the output interface of the dome.  This effect
also ensures that the apparent diameter of the emitter as is seen from
outside of the dome never exceeds the diameter of the dome, which sets
an obvious upper bound for the total gain in this structure when
$R\gg\lambda$: $G < R^2/r^2$, i.e., the whole structure may not
radiate more than a BB with radius equal to the outer radius of the
dome.

Because realistic thermal sources have $e_s<1$, and because the known
transparent materials in the infrared range have rather small
refractive indices $n\lesssim3$, the thermal lens \cite{4} can hardly
offer gain $G_{\rm BB}$ which would exceed $7$.  Here, $G_{\rm BB}$ is
the ratio of power emitted by a realistic thermal source covered with
a transparent dome to the power emitted by an uncovered BB with the
same size and temperature as the original source. In \cite{4} the gain
$G_{\rm BB}\approx 3.1$ has been experimentally demonstrated for a
thermal lens of centimeter size with $n\approx 2.4$. The above
estimations predict $G_{\rm BB} \approx 4.8$ for this case, when the
emitter is an ideal BB.

The study that we are going to present next has been motivated by the
following question: Since it is possible to enhance the thermal
radiation of an emitter by 2--5 times by using a hemisphere of a
transparent isotropic dielectric, can we go further using more
advanced materials? Namely, can we approach the GO bound: $G_{\rm max}
= R^2/r^2$ with these materials? Note that here we are interested in
the case when $R\approx 3\lambda$ or greater, because bodies with
$R\lesssim\lambda$ can outperform this bound~\cite{3}.  We show that a
dome made of a hyperbolic metamaterial theoretically allows one to
increase the spectral radiance of small emitters by up to two orders
of magnitude, as compared to the limit dictated by Planck's law for BB
emitters of the same size in vacuum. Hyperbolic metamaterials (HMM)
which we propose for this purpose are uniaxial dielectric composites
with the permittivity tensor defined by two components: transverse
$\varepsilon_{\bot}$ and axial $\varepsilon_{\|}$, such that ${\rm
  Re}(\varepsilon_{\|}){\rm Re}(\varepsilon_{\bot})<0, \ {\rm
  Re}(\varepsilon_{\|,\bot})\gg {\rm Im}(\varepsilon_{\|,\bot})$. The
isofrequency surfaces (also called wave-dispersion surfaces) for HMM
represent hyperboloids. A hot unit volume inside a HMM sample emits
much more electromagnetic energy than a unit volume of a conventional
lossy medium at the same temperature. This effect results from high
Purcell's factor of a dipole located inside HMM. The concept of
Purcell's factor (the gain in the spontaneous emission rate)
historically referred to the case when the dipole radiation was
enhanced by a closely located resonator (see e.g. in
\cite{Hecht}). However, in work \cite{Who} the notion of Purcell's
factor was extended to any environment of the dipole source different
from free space. Purcell's factor of HMM dramatically exceeds the
Purcell's factor $F_{\rm P, diel} = n$ of a usual dielectric. In the
lossless HMM without internal granularity the radiation resistance of
a point dipole oriented orthogonally to the optical axis tends to
infinity~\cite{Felsen,5} because all the power is irradiated in the
form of propagating waves. Therefore the thermal radiation of a unit
hot volume in such an ideal HMM should be infinite. For realistic
(lossy and internally granular) HMM the thermal radiation of a unit
hot volume is finite but strongly super-Planckian as compared to
vacuum \cite{1,2}.

The excessive super-Planckian radiative heat in HMM is contained in
the modes with high transverse wavenumbers $q = 2\pi/\Lambda_{\rm tr}
\ge \omega/c$. In flat uniaxial HMM slabs with the optical axis oriented
orthogonally to the interface plane these modes experience TIR at the
interface with free space and, thus, are confined inside the HMM (note
that the coupling of such modes with free space can be carried out in
asymmetric HMM \cite{ScRep,graphene,SP}, where the optical axis is
tilted to the slab interface).  As a result, the thermal radiation
from such slabs into free space does not exceed the BB limit. However,
it can be very close to it, and it is known that a half-space of HMM
mimics the BB \cite{1111,1,2}.

On the other hand, in locally uniaxial radially symmetric HMM samples
the eigenmodes which are characterized with high local wavenumbers
$q(r) \gg \omega/c$ close to the center of the sample, may attain $q(R) <
\omega/c$ at enough large radial distance $R \gg r$, because in these
modes, roughly, $q(r) \propto 1/r$. Therefore, in radially symmetric
HMM these modes can couple to the free space propagating waves if the
radius $R$ is large enough. This effect is known as
hyperlensing~\cite{hyper1,hyper2,hyper3}. Hyperlenses (HLs) were
previously designed for obtaining magnified images of subwavelength
objects.

In fact, HL is also a matching device for the radiation propagating
from its central part to free space \cite{hyper4}. Here, we suggest to
use a dome of radially symmetric HMM which operates as an infrared HL
to extract the excessive super-Planckian heat otherwise confined
within emitter's near field in the modes with high transverse
wavenumbers.  An implementation of such HMM in the infrared range is,
for example, an optically dense array of aligned metal nanowires
called wire medium (WM). WM is a spatially dispersive implementation
of HMM \cite{Simovski}. The spatial dispersion of HMM for our purpose
is not a harmful factor. On the contrary, in accordance to our
estimations the spatial dispersion helps to match the hyperlens to
free space.

Performing our emitter as a lossy dielectric body placed inside a
transparent dielectric dome both comprising radially divergent
nanowires, we arrive at the structure sketched in
Fig.~\ref{fig1}~(right half). For better matching of the HL to free
space the ends of the nanowires can be made free-standing as it is
shown in the figure.  To prevent direct thermal contact, the nanowires
in the emitter may be separated from the HL by a sufficiently narrow
nanogap.  It is critical that the nanowires are radially oriented in
the whole structure and that their density decreases with radial
coordinate. The divergence angle $\phi$ between adjacent nanowires
should be small enough so that the properties of HMM in the emitter
volume are preserved, however, large enough so that the best possible
matching to free space is achieved.

Formulas for the effective permittivity of WM operating at infrared
can be found in \cite{Simovski,Mario}.  We use highly radially
anisotropic HMM, in which ${\rm Re} (\varepsilon_{\bot})>0,\ {\rm Re}
(\varepsilon_{\|})<0$, $|\varepsilon_{\|}|\gg|\varepsilon_{\bot}|$, and the energy propagates
roughly in the radial direction, independently on the value of
$q$. Thus, we notice that in this regard the situation is similar to
the case of the simple dielectric dome considered previously, with a
difference that when estimating the power transmittance through the
outer interface of the HL we must use the effective complex index of
refraction in the WM in the vicinity of the outer interface: $n_{\rm
  \bot,out} = \sqrt{\varepsilon_{\rm\bot,out}}$. Hence, $t_{\rm HL} = 4{\rm
  Re}(n_{\rm\bot,out})/|n_{\rm\bot, out}+1|^2$.

Note that still the modes with $q(R) \ge \omega/c$ experience TIR at the
dome-air interface and for these modes $t_{\rm HL} = 0$. From recent
studies of the dipole radiation in WM \cite{Nefedov,Lovat} it is known
that the irradiated wave beam is nearly as narrow as the WM period
$a$. Therefore, in order to obtain high transmittance to free space
(nearly as high as $t_{\rm HL}$) for the dominant part of the spatial
spectrum exited within HL, the divergence angle $\phi$ should be such
that the separation between the nanowires at the outer surface of the
HL is about $\lambda/2$ or larger. Hence, $\phi \gtrsim \lambda/(2R)$.

Let us now estimate how large can be the gain $G_{\rm HL}$ in the HL
configuration of Fig.~\ref{fig1}.  First, we note that inclusion of
nanowires into a dielectric host increases the power radiated by an
elementary dipole placed inside this medium by $F_{\rm P}$ times,
where $F_{\rm P}$ is the Purcell factor for uniaxial WM. This factor
was calculated in Ref.~\cite{5}. Being averaged over all possible
locations of a transversely oriented electric dipole, $F_{\rm P}$
equals
\begin{equation}\label{eq:formula} F_{\rm P}^{\rm tr}\approx {3k_{\rm p}^2 \over
  8k_{\rm h}^2} \log\left[1+{K_m^2\over k_{\rm p}^2}\right],
\end{equation}
where
$k_{\rm h} = \sqrt{\varepsilon_{\rm h}}\omega/c$ and $k_{\rm p} =
\sqrt{2\pi/\log[a^2/4r_0(a-r_0)]}/a$ are the wavenumber in the host
medium and the plasma wavenumber in WM~\cite{StasPRB}, respectively,
where $a$ is the WM period and $r_0$ is the wire radius. In
Eq.~(\ref{eq:formula}), $K_m$ is the spatial spectrum cut-off parameter,
which equals $2\sqrt{\pi}/a$ in unbounded uniaxial
WM~\cite{5}. Because only the modes with $q < \omega/c$ are irradiated
from HL's outer surface, here we must limit this parameter by $K_m =
\min[2\sqrt{\pi}/a, (\omega/c)(R/r)]$. Strictly speaking, Eq.~(\ref{eq:formula})
refers to the case when the wires are perfectly conducting, however in
\cite{5} the estimations were done also for lossy wires and it was
shown that (\ref{eq:formula}) was applicable to realistic metal nanowires if
they were thick enough (practically, their radius $r_0$ should be
larger than the skin depth).  For dipoles parallel to the wires the
Purcell factor is much smaller than $F_{\rm P}^{\rm tr}$ and can be
neglected. Since a hot elementary volume of a lossy medium surrounded
by nanowires can be treated as a set of three identical mutually
orthogonal dipoles emitting thermal radiation, in thermal emission
calculations we must use the average $F_{\rm P}^{\rm HL} = 2F_P^{\rm
  tr}/3$, where $F_P^{\rm tr}$ is given by (\ref{eq:formula}).

Under these conditions, the total gain due to the effect of the HL
dome can be estimated as follows:
\begin{equation}
G_{\rm HL} \approx F_{\rm P}^{\rm HL}\times{\rm Re}(n_{\rm\bot,in})^2\times t_{\rm
  HL}\times e^{-2\alpha(R-r)}
          \approx {8F_{\rm P}^{\rm tr}\times{\rm Re}(n_{\rm\bot,in})^2{\rm
              Re}(n_{\rm\bot,out})\over 3|n_{\rm\bot, out}+1|^2}e^{-2\alpha(R-r)},
\label{eq:po}
\end{equation}
where $n_{\bot,\rm in}$ is the effective refractive index of the HL in
the vicinity of the emitter (at the inner interface of the HL), and
$\alpha \approx (\omega/c){\rm Im}(\sqrt{\varepsilon_{\bot}})$ is the decay factor
due to the loss in the nanowires. Note that the structure suggested in
this paper cannot radiate more than a BB with the same size as that of
the dome and the same temperature as that of the emitter when
$R\gg\lambda$. Moreover, because the WM-based HL interacts mostly with
$P$-polarized waves, the actual upper bound for the gain in this case
is $G_{\rm HL}^{\rm max}\sim 0.5R^2/r^2$.

Due to optical losses in the metal the decay of thermal radiation over
the path $R\gg \lambda$ is not negligible. This factor restricts the
radius $R$ by dozens of $\lambda$. However, Eq.~(\ref{eq:po}) does not take
into account the thermal radiation of heated wires inside the
hyperlens. This additional emission can significantly increase $G_{\rm
  HL}$ compared to (\ref{eq:po}), so that it may approach the GO bound:
$G_{\rm max}=R^2/r^2$. In the same time, (\ref{eq:formula}) slightly
overestimates the Purcell factor for realistic nanowires. So, the
implementation of our thermal HL with macroscopic dimensions
(e.g. with $R=1$ cm like in \cite{4}) is disputable. In the present
study we deal with a microscopic hyperlens with radius $R=10\ \mu$m
and an emitter of radius $r=0.75\ \mu$m.

Estimations of the factor $G_{\rm HL}$ were done using Eq.~(\ref{eq:po}) and
the effective-medium model of infrared WM \cite{Mario}.  The material
parameters of gold were taken from Ref.~\cite{Palik}.  We considered a
hemispheric structure with concentric hemispheric emitter located on a
perfect mirror as in Fig. \ref{fig1}. The HL is performed of
non-tapered gold nanowires with the thickness $2r_0=50$~nm located in
a matrix with $\varepsilon_2=3.16$ (chalcogenide glass transparent in the range
50--150 THz). We calculate the gain $G_{\rm HL}$ in the range 100--140
THz, where $r_0>\delta$ ($\delta$ is the skin-depth of gold), and
formula (\ref{eq:formula}) for Purcell's factor is applicable. Internal ends
of nanowires are located at $r_1=0.5~\mu$m from the geometrical center
of the structure.  An emitter comprises the hemisphere $r=0.75~\mu$m
and is partially filled with nanowires. The emitter is assumed to be a
lossy dielectric which is well impedance-matched with the HL. The
distance between the centers of nanowires at the surface
$r_1=0.5~\mu$m equals $100$~nm and within the emitter the averaged
period of the WM equals $a=125$ nm. Nanowires diverge with the angle
$\phi\approx 10^{\circ}$. This angle is small enough to neglect the
divergence of nanowires when calculating the effective permittivity of
HMM in the domain of the emitter and its Purcell factor $F_{\rm
  P}^{\rm tr}$. However, it is large enough to offer good matching of
the HL to free space, because for $\phi \gtrsim 7.5^{\circ}$ the
distance $A$ between the axes of nanowires at the outer surface of the
HL exceeds $\lambda/2$ at frequencies 100--140 THz.

Following to (\ref{eq:formula}) Purcell's factor of the medium of parallel
nanowires with the period $a=125$ nm for a transverse electric dipole
decreases from $F_{\rm P}^{\rm tr}\approx18$ to $F_{\rm P}^{\rm
  tr}\approx7.5$ over the range 100--140~THz.  Then the relative
enhancement $G_{\rm HL}$ of the power spectrum radiated by an
arbitrary dipole $\_p$ located in between the wires near the internal
surface of the HL in accordance to (\ref{eq:po}) is within approximately
$40\dots 20$ over this frequency range. The range 100--140 THz is
around the maximum of emission for the emitter temperatures $T_1$ of
the order 700--800$^{\circ}$C. Higher temperatures are hardly actual
for our HL since nanowires can melt. For lower temperatures thermal
radiation is concentrated in lower frequencies where Purcell's factor
is higher. For example, at 50 THz in unbounded WM $F_P^{\rm tr}\approx
70$. The same divergence angle at this frequency implies larger $R$
needed for matching the HL to free space. The condition $A>\lambda/2$
holds at 50 THz for $R=20$~$\mu$m. Then, taking into account the decay
we obtain using (\ref{eq:po}) the gain $G_{\rm HL}\approx 170$ at 50 THz. So,
for emitters with temperatures $T_1<$500--600$^{\circ}$C the thermal
radiation of the emitter within HL may exceed the BB limit for the
same emitter in vacuum by two orders of magnitude.

To check our estimations of the gain $G_{\rm HL}$ we performed extensive
numerical simulations. We studied a HL excited by a transverse dipole
located in the middle between the ends of adjacent nanowires either at
the surface of the central nanocavity or displaced from this surface
-- either embedded into the WM up (to 250 nm from the central cavity)
or located inside it. The parameters of the HL in these simulations
are as above besides one replacement -- we substituted gold nanowires
by perfectly conducting ones. This replacement dramatically reduced
the computation time needed for the structure comprising many hundreds
of metal nanowires and made simulations realistic. Simulations were
performed using the CST Studio Suit software.

Although replacing gold by perfect conductor we removed the decay
factor $\exp(-2\alpha R)$, this is still a reasonable model of a
HL. The decay factor is not the most relevant parameter and can be
easily taken into account analytically. The absence of absorption
makes the relative enhancement of radiation into free space equivalent
to Purcell's factor. This equivalence allows us to concentrate on the
hyperlensing effect, i.e., on emission enhancement and matching of our
structure to free space. Our model source is a very short dipole
antenna of perfectly conducting wire with bulbs mimicking the Hertzian
dipole at the simulation frequency.

\begin{figure}[h!]
\centering \epsfig{file=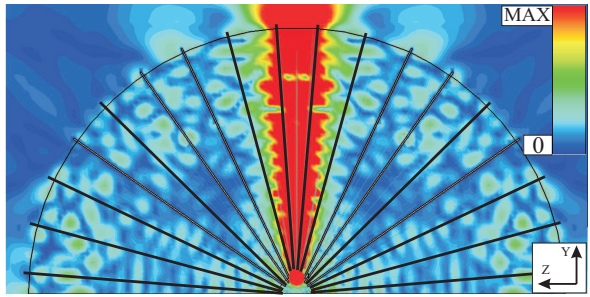, width=0.45\textwidth}
\caption{Electric field amplitude distribution in the $H$-plane
  (vertical cross section orthogonal to the dipole) produced by a
  dipole located in between the internal ends of perfect nanowires of
  radius $r_0=25$ nm forming our HL. Its host material (between
  $r_1=0.5\ \mu$m and $R=10\ \mu$m) is glass.} \label{fig2}
\end{figure}

\begin{figure}[h!]
\centering \epsfig{file=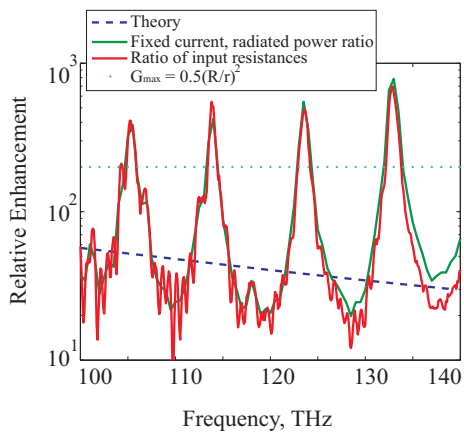, width=0.35\textwidth}
\caption{Relative enhancement of radiation by a transverse dipole due
  to the presence of a HL of perfect wires calculated 1) directly via
  the radiated power spectrum (green curve) and 2) through the input
  resistance of a short wire dipole (red curve).  The structure is the
  same as in Fig.~\ref{fig2}.  The theoretical blue dashed curve and
  the green dotted line are explained in the main text.
} \label{fig3}
\end{figure}

First, we calculate the field distributions to inspect if the wave
beam divergence is sufficient to prevent strong reflections from the
effective surface of the HL. For divergence angles within the range
$\phi=6$--$9^{\circ}$ the concept of HL turned out to be fully
adequate, and the result weakly depends on $\phi$ over this interval
of values. The result weakly depends on the exact location of the
transverse dipole embedded into the WM up to 250 nm from the internal
cavity $r_1$. However, if the dipole is moved to the central cavity to
the distance more than 250 nm, Purcell's factor drops to unity. Also,
the radiation decreases if $\phi<6^{\circ}$ i.e. when the HL
approaches to a block of parallel nanowires. In Fig.~\ref{fig2} a
color map illustrates the hyperlensing of the dipole radiation for
optimal divergence angle $\phi=10^{\circ}$.  The horizontal dipole is
located on top of the internal cavity $r_1$ in between two central
nanowires and radiates at the frequency 120 THz which is between the
bands of Fabry-Perot resonances. In both $E$- and $H$-planes we
observed a sufficient width of the main radiation beam. The reflection
from the effective surface of the HL in our simulations fits the
estimation $t\approx t_{\rm HL}$.

For a lossless HL $G_{\rm HL}$ can be calculated in two ways: via the
input resistance of the antenna and via the far-zone radiated
power. Both these values were calculated and normalized to the
corresponding values simulated for the same dipole when the HL is
absent. In the first case we keep the same input voltage of the
antenna in the absence or presence of the HL. In the second case we
fix the antenna current. The coincidence of two results is expected at
low frequencies where the short wire antenna is close to the Hertzian
dipole.  This equivalence is seen in Fig.~\ref{fig3} at 100--130 THz
where the red and green curves nearly coincide (besides the small
ripples of the red curve which are numeric errors). In this plot we
observe several Fabry-Perot resonances at which the HL gain reaches
very high values. These values are, however, hardly relevant for the
thermal radiation because the emitter mimicking the BB will absorb all
incoming waves.  Therefore, the Fabry-Perot resonances in the HL in a
more realistic configuration will be greatly suppressed. The blue curve
shows the theoretical estimation for the HL gain calculated in
accordance to Eq.~(\ref{eq:po}) with the factor $(2/3)\exp(-2\alpha R)$
excluded because only single orientation of the dipole in a lossless
HL is considered in the simulations. In the range 100--130 THz our
theoretical estimation agrees with the simulated gain when averaged
between the Fabry-Perot resonances.

To conclude, we have suggested a structure that greatly enhances the
radiative heat power produced by a small thermal emitter, which may go
far beyond the limit enforced by Planck's law for the same radiator in
free space. This is achieved by centering the emitter at the focal
point of a hyperlens, which transforms emitter's near field into
propagating waves which are matched well to free space and efficiently
irradiated. However, the structure suggested in this paper still
radiates less than a BB with the same size as that of the hyperlens
and the same temperature as that of the emitter. A theoretical
possibility to overcome this restriction for bodies of constrained
radius is reserved for a future work (see \cite{arxivNanorad}).

\end{document}